\documentclass{appolb}
\usepackage{epsfig}
\newcommand{\lb}[1]{#1\!\!\!^-} 

\begin{document}
\title{On Theoretical Problems in Synthesis of Superheavy Elements
}
\author{Yasuhisa  ABE
\and
Bertrand BOURIQUET
\address{Yukawa Institute for Theoretical Physics
               Kyoto Univ., Kyoto 606-8502, JAPAN}
}
\maketitle
\begin{abstract}
Towards precise predictions of residue cross sections of the
superheavy elements, recent theoretical developments of reaction
mechanisms are presented, together with the remaining problems which
give rise to ambiguities in absolute values of cross sections predicted.
\end{abstract}
  
\section{Introduction}

   Predictions and structure studies of superheavy elements  (SHE) have been made,
 since the establishment of the nuclear shell model\cite{bib1}.   Especially in the 
last decade, elaborate investigations have been performed of shell correction 
energy and thereby of a possible location of the superheavy island in the 
nuclear chart.  Furthermore, not only about the center of the island, but also
about stability properties of nearby nuclei have been being investigated, 
which is useful for the extension of the nuclear chart in heavy and superheavy 
elements\cite{bib2}.   On the other hand, studies of nuclear reaction mechanisms 
have not been developed so much, though so-called fusion-hindrance was
 experimentally found 
to exist in heavy ion fusions and inferred to be due
 to energy 
dissipation\cite{bib3}.   That is, there is no reliable theoretical framework which 
enables us to predict fusion probability of massive systems and
 thereby residue cross sections 
 of SHE properly.   Thus, which combination of incident ions is most
 promising and 
what incident energy is an optimum is not yet predicted theoretically.   
Therefore, the fusion experiments have been performed, based on systematics of 
data available so far\cite{bib4}.  

Based on the reaction theory of the compound nucleus\cite{bib5}, 
residue cross sections are given as follows, 

\begin{equation}
\sigma=\pi{\lb{\lambda}}^2\Sigma(2J+1)\cdot P^J_{\mbox{\scriptsize
fusion}}\cdot P^J_{\mbox{\scriptsize surv}},
\end{equation}

\noindent
where $\lb{\lambda}$ is the wave length divided by $2\cdot\pi$ and
$J$ the total angular momentum of the system.
$P_{\mbox{\scriptsize fusion}}$ and $P_{\mbox{\scriptsize surv}}$ are 
fusion and survival probabilities, respectively.  In the present 
paper, we discuss several difficult problems inherent in synthesis of the 
superheavy elements with brief explanations of a few progresses of our 
understanding, as well as attempts of realistic calculations.

\section{Difficulties Characteristic in Synthesis of SHE}

   In order to obtain the fusion probability, we have to take into
   account possible
mechanisms for the fusion-hindrance.   Otherwise, calculated
probabilities, and 
fusion cross sections would be unrealistic, as it is the case that one uses a 
transmission coefficient of an optical model or a barrier penetration factor 
as the fusion probability.  As for possible origins of the hindrance,
two mechanisms
 are proposed.  One is dissipation of incident energy in the 
course of two-body collisions and thus probability for the system to overcome 
the Coulomb barrier is reduced\cite{bib7}.  
The other one is dissipation of energy of 
collective motions of the amalgamated system which has to overcome a 
conditional saddle or a ridge line in order to reach the spherical shape, i.e.,
the compound nucleus\cite{bib8}.  Thus, the probability for reaching the spherical shape 
is also reduced.   
It is natural to consider that both exist.
In other words, the fusion probability $P_{\mbox{\scriptsize fusion}}$
consists of two 
factors; the sticking probability $P_{\mbox{\scriptsize stick}}$ of two incident ions after overcoming 
the Coulomb barrier  and the formation probability $P_{\mbox{\scriptsize form}}$ of the spherical shape 
after overcoming the conditional saddle point, starting from a pear-shaped 
configuration made by the sticking of the incident ions\cite{bib9}.  

\begin{equation}
P^J_{\mbox{\scriptsize fusion}}=P^J_{\mbox{\scriptsize stick}}\cdot P^J_{\mbox{\scriptsize form}}.
\end{equation}

   Since the existence of the saddle point or the ridge line between a pear-
shape made by the incident ions and the spherical shape is typical in very 
heavy systems, the latter mechanism would be indispensable for the fusion-
hindrance observed in massive systems, though the former would also play a 
role (Note that in lighter heavy-ion systems the amalgamated shape is usually 
located inside the ridge line, so the system eventually slides down to the 
spherical shape with probability being equal to 1, once the incident ions 
stick to each other, though fluctuations to be discussed below may reduce it 
only slightly).  
   In either mechanism, we have to describe a passing over a barrier under 
energy dissipation, which is not yet well understood theoretically\cite{bib10} and 
thereby there is no useful formula ready for practical applications.   This is 
quite contrast to a similar problem, i.e., to fission under dissipation, where 
a famous Kramers formula\cite{bib11} for decay rate is known to well describe a process of 
a system inside a potential pocket leaking over the fission barrier.   An 
essential difference is that in the latter the initial system is in
the 
quasi-equilibrium in the pocket, while in the former the initial state is given by 
the condition of two incident ions with a given incident c.m.energy.   

   Recently, the present author and his collaborators have proposed a new 
analytic formula for the probability of passing over a parabolic barrier under 
frictional force.   We have applied this formula to the problem of passing-%
over a conditional saddle point, and obtained a simple expression for
so-called 
extra-push energy which provides a clear understanding of the fusion
hindrance\cite{bib12,bib13}.  This 
would be an important contribution to study of fusion mechanisms and is 
briefly recapitulated in section 3., but there still remains a difficulty in 
practice. 
   The parabolic shape is usually a good approximation for barrier shapes, but 
in potential landscapes calculated with the liquid drop model (LDM) a pocket 
inside the saddle is very shallow in nuclei corresponding to the superheavy 
elements, as is easily expected from the fissility parameter $x_f$ being close 
to 1.  Therefore, the potential is expected to be substantially asymmetric 
around the saddle, and moreover a system once passing over the saddle may 
return back with an appreciable probability due to the fluctuation
associated with the friction.   Of course, the probability for 
return-back to re-separation is reduced if the system is cooled down by 
neutron emissions and restores the shell correction energy which makes the 
pocket deeper.  

For a quantitative prediction, those features should be taken 
into account properly, which is made by numerically solving a Langevin 
equation\cite{bib14}.   
For a dynamical description of shape evolutions, we have to solve
trajectories in a multi-dimensional space of shape parameterization with a
realistic LDM potential, examples of 
which are discussed in section 5.   But a time-dependent shell correction 
energy due to evaporation of neutrons is not yet taken into account in fusion 
processes. (Since time for fusion process is expected to be rather short, this 
would not cause a serious inaccuracy, but is properly done in the calculation 
of survival probability, as will be discussed in sections 6. and 7.) 
   In the approaching phase of passing over the Coulomb barrier under
friction, we have  
 to take into account a dissipation of the orbital angular momentum as well 
as that of the kinetic energy of the radial motion, where a coupling
between them is 
not described by a quadratic potential, as is discussed in section 4. 
Of course, there are many other effects which may play a role in the latter 
process, say, effects of deformations of incident ions, quantum
tunneling effects etc., which are not yet fully investigated.   

   As for the survival probability, the statistical theory of decay is well 
established for obtaining a probability for the system to survive against 
fission and charged particle decay.   But in practice there are ambiguities 
in the physical parameters, i.e., so-called level-density parameter $a$ and the 
shell damping energy $E_d$ which controls restoration of the shell correction 
energy by cooling.   Especially, the latter is crucially important, because 
the restoring shell correction energy gives rise to an additional fission 
barrier effectively which controls survival probability, which is
discussed qualitatively in section 6. and quantitatively in section 7.
In addition, there 
are Kramers\cite{bib11,bib15} and collective enhancement
factors\cite{bib16} in fission decay widths to be 
taken into account.  These are briefly discussed and examples of
realistic calculations on $^{48}$Ca+ actinide targets are presented
which are made by employing a new statistical code KEWPIE\cite{bib17} for the
survival probabilities, in section 6.   

\section{Fusion Hindrance and Extra-Push Energy :  Parabolic Barrier}

   We study a problem of obtaining a probability for passing over a
potential barrier 
under a frictional force, which originates from interactions of the degree 
of freedom under investigation with a heat bath, i.e., with other degrees of 
freedom.  Therefore, there should be a random force associated with the 
friction in accord with the dissipation-fluctuation theorem.   If we 
approximate the barrier with an inverted parabolic shape, the equation
of motion for a  
coordinate $q$ and its associate momentum $p$ is written as follows,

\begin{equation}
\frac{d}{dt}\left(
\begin{array}{c}
q\\
p
\end{array}
\right)=
\left(
\begin{array}{cc}
0&1/m\\
m\omega^2&-\beta
\end{array}
\right)
\left(
\begin{array}{c}
q\\
p
\end{array}
\right)+
\left(
\begin{array}{c}
0\\
R
\end{array}
\right),
\end{equation}

\noindent
where $m$ denotes the inertia mass,and $\omega$ the curvature of the inverted 
parabola.  $\beta$ is a reduced friction, i.e., the friction $\gamma$
divided by $m$, 
while $R$ is its associated random force.
The random force is assumed to be Gaussian and satisfies the flowing properties, 

\begin{eqnarray}
\left<R(t)\right>&=&0,\\\nonumber
\left<R(t)\cdot R(t')\right>&=&2\cdot\gamma\cdot T\cdot\delta(t-t'),
\end{eqnarray}

\noindent
where $\left<\right>$ signifies an average over all the possible
realizations 
and the last equation given in Eq.~(4)
with temperature $T$ of the heat bath 
expresses the dissipation-fluctuation theorem.   Since the equation is
linear, one can write down a 
general solution.   

With this solution an general expression for a 
distribution function $W(q,p;t)$ at any later time t is calculated, starting 
with the following definition,

\begin{equation}
W(q,p;t)=\left<\delta\left(q-q(t)\right)\cdot\delta\left(p-p(t)\right)\right>_{\left\{R\right\}},
\end{equation}

\noindent
where $q(t)$ and $p(t)$ denote a general solution of Eq.~(3)
and is given by 
a linear combination of their initial value $q_0$ and $p_0$ with the coefficients 
including parameters $\beta$ and $\omega$.   
$\left<\right>_{\left\{R\right\}}$ again denotes the average over all the
possible realizations of $R(t)$.
Using the path integral technique, we 
can perform the averaging and obtain the distribution function of the 
system as a Gaussian distribution around the mean trajectory $\left(\left<q(t)\right>,\left<p(t)\right>\right)$\cite{bib12}.
Then, a probability for passing over the barrier is calculated by integrating 
over the whole $p$-space and the half $q$-space, and then by taking the
limit of time $t$ to the infinity,

\begin{equation}
P_{\mbox{\scriptsize form}}=\frac12\mbox{erfc}\left[\sqrt{\frac{\sqrt{x^2+1}+x}{2x}}\left\{\sqrt{\frac{B}{T}}-\frac{1}{\sqrt{x^2+1}+x}\sqrt{\frac{K}{T}}\right\}\right],
\end{equation}

\noindent
where $x$ denotes $\beta$ divided by $2\cdot\omega$.   $K$ and $B$ denote the initial 
kinetic energy $p_0^2/2\mu$ and the barrier height measured from the initial 
potential energy  $\mu\cdot\omega^2\cdot q_0^2/2$, respectively.  

   In order for the probability to be 1/2, the argument of the error function 
should be equal to zero, which means that the mean trajectory just reaches at 
the top of the barrier overcoming the friction.   Then, the necessary critical 
kinetic energy $K_c$ is given as 

\begin{equation}
K_c=\left(\sqrt{x^2+1}+x\right)^2\cdot B,
\end{equation}

\noindent
where we can see that in the case of no friction, i.e., of $x$ being equal to 
zero, $K_c=B$, which is trivial.   It clearly shows that $K_c$ is much larger than $B$ 
under the frictional force.  If we estimate the first factor in Eq.~(7)%
,
assuming One-Body Wall-and-Window formula (OBM\cite{bib18}) for the friction $\gamma$, we 
obtain about 10, depending on a reasonable choice of values for the 
inertia mass and the curvature of the potential calculated with LDM.  This 
gives a simple formula for the extra-push energy, though we should be careful 
in a comparison with experimental data about effective one-dimensional 
quantities for $B$, $\mu$, $\omega$ and Coulomb barrier heights of 
entrance channel etc.    

   Another interesting formula is obtained, which is very useful for synthesis 
of SHE.   Residue cross sections are extremely small in SHE, i.e., we are 
facing with the situation where fusion probability is very small.  This 
suggests in our present formulation that the mean trajectory does not reach 
the top of the barrier, even is far before the top, which means that the 
argument of the error function of Eq.~(6)
is very large.  Then, employing 
an asymptotic expansion of the error function, we can obtain a simple 
approximate formula for the formation probability $P_{\mbox{\scriptsize form}}$\cite{bib13},

\begin{eqnarray}
P_{\mbox{\scriptsize
form}}\cong\frac12\frac{1}{\sqrt{\pi}}\left[\sqrt{\frac{\sqrt{x^2+1}+x}{2x}}\left\{\sqrt{\frac{B}{T}}-\frac{1}{\sqrt{x^2+1}+x}\sqrt{\frac{K}{T}}\right\}\right]^{-1}\\\nonumber
\cdot\exp\left[\sqrt{\frac{\sqrt{x^2+1}+x}{2x}}\left\{\sqrt{\frac{B}{T}}-\frac{1}{\sqrt{x^2+1}+x}\sqrt{\frac{K}{T}}\right\}\right],
\end{eqnarray}

\noindent
where it is interesting to note that there is a factor very similar to 
Arrhenius factor which is typical in thermal activation processes such as 
nuclear fission, neutron evaporation, thermal electron emission from metal,
etc.   The exact Arrhenius factor is obtained in case of a complete 
damping of the relative motion in the approaching phase, as follows.   As will 
be shown in the next section, a distribution of the radial momentum at the 
contact point is approximately expressed by a Gaussian as a results of two-body collision processes and thus, the formation probability is obtained by a 
convolution over initial momentum $p_0$, which results again in an error function 
of Eq.~(6)
with K being replaced with the average value $\bar K$ and with the 
associated variance.   In case of completely damping, $\bar K$ is equal to zero, 
and the variance is equal to the temperature.  Accordingly, the corresponding 
asymptotic expansion gives the exact Arrhenius factor.
   
\begin{equation}
P_{\mbox{\scriptsize
form}}\cong\frac12\frac{1}{\sqrt{\pi}}\sqrt{\frac{T}{B}}\exp\left[-\frac{B}{T}\right]
\end{equation}

   Since fusion is inverse to fission in reaction directions, one could call 
this as an ``inverse Kramers formula\cite{bib19}''.   But we should be careful that in the 
thermal activation processes the factor appears in decay rate or in emission 
rate per unit time, while in the present case it appears in the transition 
probability, i.e., time-integrated quantity.  Anyhow, a physical meaning of 
the factor as well as of the pre-exponential factor are yet to be understood.  

   For actual fusion reactions, one-dimensional treatment is obviously an 
over-simplification, which is readily understood by considering a 
mass-asymmetric entrance channel.   
In addition, neck formation etc. would come 
into play.     It is worth to notice that even in such situations, i.e., in 
multi-dimensional problems, we can derive the same type of formula as 
Eq.~(6)%
, starting with an assumption of a quadratic potential generalized 
to a multi-dimension.    That indicates that we can define an
effective one-dimensional model.   
Thus, the qualitative 
understandings obtained above with the schematic one-dimensional model are 
considered to be useful, with the barrier height etc. being considered 
to be effective quantities. 

\section{Approaching Phase ; Passing-Over Coulomb Barrier under Friction}

   One could apply the formula obtained in the previous section to 
passing-over Coulomb barrier, approximating again the barrier as an inverted 
parabola.  There, however, is another problem, as stated in section 2.   In 
the approaching phase,dissipation of the orbital angular momentum comes into 
play, coupled with the radial motion.   For the problem, the most simple 
and readily applicable model is the surface friction model (SFM), proposed 
by Gross and Kalinowski\cite{bib6} in order to explain so-called Deep-Inelastic 
Collisions.  

   Below, we reformulate it, starting with a general framework of Ref.~[20] which 
includes so-called rolling friction.    Starting with intrinsic spins of the 
incident ions, $L_1$ and $L_2$, respectively, we introduce the
following variables,
             
\begin{eqnarray}
L^+&=&L_1+L_2=L_0-L(t)\\\nonumber
L^-&=&\left(C_1L_2-C_2L_1\right)/C_1+C_2,
\end{eqnarray}

\noindent
where $L_0$ denotes an incident orbital angular momentum and $C_i$,
$i$ being 1 or 2, is an effective ion radius defined as follows,

\begin{equation}
C_i=R_i\left(1-\left(b/R_i\right)^2\right),
\end{equation}

\noindent
where $b=1\mbox{fm}$ and $R_i=1.28\cdot A^{1/3}_i-0.76+0.8\cdot
A^{-1/3}_i$ with $A_i$ being the mass number of $i$-th ion.
Then, a Langevin equation for two-body collisions is written as 

\begin{eqnarray}
\frac{dr}{dt}&=&\frac{1}{\mu}p\\
\frac{dp}{dt}&=&-\frac{dV}{dr}-\beta_r\cdot p+\theta_r\cdot\omega_r(t)\\
\frac{d}{dt}
\left(
\begin{array}{@{}c@{}}
L\\
L^-
\end{array}
\right)&=&
\left(
\begin{array}{@{}cc@{}}
\beta_{11}&\beta_{12}\\
\beta_{21}&\beta_{22}
\end{array}
\right)
\left(
\begin{array}{@{}c@{}}
L\\
L^-
\end{array}
\right)+
\left(
\begin{array}{@{}c@{}}
\tilde{\beta_{11}}\\
\beta_{21}
\end{array}
\right)\cdot L_0+
\left(
\begin{array}{@{}cc@{}}
\theta_{11}&\theta_{12}\\
\theta_{21}&\theta_{22}
\end{array}
\right)
\left(
\begin{array}{@{}c@{}}
\omega_1\\
\omega_2
\end{array}
\right),
\end{eqnarray}

\noindent
where $\mu$ is equal to the reduced mass of the entrance channel, and $V$ 
denotes a sum of the Coulomb $V_c$ and the nuclear $V_n$ potentials with the 
rotational energy given by the orbital angular momentum $L$.  The friction 
tensor $\beta_{ij}$ and $\beta_r$ are given below,

\begin{eqnarray}
\beta_r&=&C_r\Psi(r)/\mu\\\nonumber
\left(
\begin{array}{@{}c@{\,}c@{}}
\beta_{11}&\beta_{12}\\
\beta_{21}&\beta_{22}
\end{array}
\right)&=&
\left(
\begin{array}{@{}c@{\,}c@{}}
-\Psi(r)\cdot
C_T\cdot\left[\frac{1}{\mu}+\frac{r^2}{\left(C_1+C_2\right)^2}\cdot\frac{C^2_1J_2+C^2_2J_1}{J_1\cdot 
J_2}\right]&-\Psi(r)\cdot
C_T\left[\frac{r^2}{C_1+C_2}\cdot\frac{C_1J_2-C_2J_1}{J_1\cdot
J_2}\right]\\
-\Psi(r)\cdot
C_{\mbox{\scriptsize roll}}\cdot g^2\left[\frac{1}{C_1+C_2}\frac{C_1J_2-C_2J_1}{J_1\cdot
J_2}\right]&-\Psi(r)\cdot
C_{\mbox{\scriptsize roll}}\cdot g^2\cdot\frac{J_1+J_2}{J_1\cdot J_2}
\end{array}
\right),
\end{eqnarray}

\noindent
where $\Psi(r)$ is a form factor specified below, and $C_T$ and $C_{\mbox{\scriptsize roll}}$ denote strengths for 
tangential and rolling frictions, respectively,    In addition, a parameter 
$g$ is introduced for describing a effective depth of the rolling friction, 
which is taken to be 1.0fm.   $J_i$, $i$ being 1 and 2, are the rigid moment of 
inertia of the incident ions which are assumed to be spherical.   
Then, the strengths $\theta_{r}$ and $\theta_{ij}$ are adjusted to
satisfy the dissipation-%
fluctuation theorem with the friction tensor $\beta_r$ and $\beta_{ij}$.   
The coefficient $\tilde{\beta}_{11}$ is given by 
$\beta_{11}-\Psi(r)\cdot C_T/\mu$.  
Langevin forces are given by $\omega_i$, $i$ being $r$, 1 and 2 which denote
Gaussian 
random numbers and are assumed to have the following properties,

\begin{equation}
\left<\omega_i\right>=0,
\left<\omega_i(t)\omega_j(t')\right>=2\cdot\delta_{ij}\cdot\delta(t-t')
\end{equation}

If one 
wants to introduce deformations of the ions, one has to introduce additional 
degrees of freedom which describe their orientations.   
   If we assume that the rolling friction is very weak compared with the 
others, we take $C_{\mbox{\scriptsize roll}}$ to be zero.  Then, $d
L-/dt=0$, and $L_-=\mbox{constant}=L_-(-\infty)=0$.  The 
equation for the orbital angular momentum is rewritten simply as follows,

\begin{equation}
\frac{dL}{dt}=-K_{\Phi}/\mu\cdot\left(L-L_{st}\right)+\theta_{11}\cdot\omega_1,
\end{equation}

\noindent
where the effective friction $K_{\phi}$ and the limiting angular momentum $L_{st}$ 
are given as 

\begin{eqnarray}
K_{\phi}&=&C_T\cdot\left[1+\frac{\mu
r^2}{(C_1+C_2)^2}\cdot\left(\frac{C^2_1}{J_1}+\frac{C^2_2}{J_2}\right)\right]\cdot\Psi(r)\\\nonumber
L_{st}&=&\left(\frac{C^2_1}{J}+\frac{C^2_2}{J}\right)\cdot\mu
r^2/\left[\left(C_1+C_2\right)^2+\left(\frac{C^2_1}{J_1}\frac{C^2_2}{J_2}\cdot\mu 
r^2\right)\right]
\end{eqnarray}

If we approximate $R_i\cong C_i$, 

\begin{eqnarray}
K_{\phi}&\cong&\frac72C_T\Psi(r),\\\nonumber
L_{st}&=&\frac57L_0.
\end{eqnarray}

\noindent
$L_{st}$ is so-called 
rolling limit, while one could obtain the sticking limit if one takes the 
limit that the drift part of the r.h.s. of Eq.~(15)
is equal to null vector, 
as discussed in Ref.~[21].  
Together with $K_r=\mu\beta_r=C_r\cdot\Psi(r)$, Eqs.~(12), (13) and (17)
just correspond to SFM.  The correspondence 
is precised by giving the following relation of the friction forces,

\begin{equation}
\Psi(r)=\left(\frac{dV_N}{d_r}\right)^2, \;\;\; K^0_{\phi}=\frac72C_T=0.01, \;\;\; K^0_r=C_r=4,
\end{equation}

\begin{figure}
\begin{center}
\includegraphics[height=10cm]{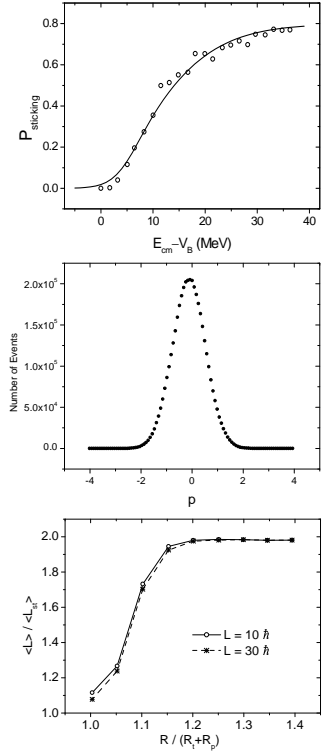}
\end{center}
\caption{Results of SFM calculations for $^{48}$Ca+$^{244}$Pu system.
The top panel shows the sticking probability $P_{\mbox{\scriptsize stick}}$ as a
function of incident energy relative to the Coulomb barrier.  
The middle panel shows the distribution of the radial momentum at the contact
point, where $P$ is given in unit of 10$^{-21}$sec$\cdot$MeV/fm.  
The bottom panel shows the orbital angular momentum divided by the
limit $L_{st}$ as a function of
relative distance.}
\end{figure}

\noindent
where the numerical values are given in unit of 10$^{-23}$s/MeV.
The dissipation-fluctuation theorem is satisfied by the equations:
$\theta^2_r=K_r\cdot T_A$ and $\theta^2_{11}=r^2\cdot K_{\phi}\cdot T_A$
with $T_A\equiv T_A(t)$ being temperature of the colliding system in
the approaching phase.
Examples of numerical solutions 
with the proximity model and 
with SFM are given in Ref.~[22].  Generally speaking, the former is much 
weaker than the latter, but for the moment we cannot draw a definite 
conclusion on which one is correct or more realistic.   The former neglects 
frictional force stemming from strong inelastic excitations etc., while the 
latter does a rolling friction.   

   We have applied SFM to superheavy systems, such as $^{48}$Ca + actinide 
targets.    
As an example, we discuss results on $^{48}$Ca + $^{244}$Pu system in detail.
The top panel of Fig.~1 shows probability $P_{\mbox{\scriptsize stick}}$ for the entrance system to reach 
the contact point, i.e., the relative distance being equal to a sum of the 
half density radii of the incident ions as a function of
$E_{\mbox{\scriptsize c.m.}}$ relative to the barrier 
height.    If there is no friction, it should be always equal to 1 above 
the barrier height (Below the barrier, it is equal to zero, since the equation 
is classical.).  But the results are not like that.  It starts with an 
extremely small value at the barrier height energy and slowly increases to 
reach 1/2 about 12 MeV above the barrier, which would explain a part of the 
extra-push.    More interesting is that a distribution of the radial momentum 
calculated at the contact point is found to be approximately of a Gaussian 
one
with its average value being almost exactly equal to zero as shown in
the middle panel of Fig.~1, which indicates that 
the relative motion is completely damped at the contact point.   In fact, the 
average orbital angular momentum also approaches to the 
dissipation limit at
the contact point, which is shown in the bottom panel of Fig.~1.
This is also the case for other 
actinide targets, say, $^{248}$Cm and $^{252}$Cf.   

In brief, the analyses of the 
approaching phase provide us with sticking probability $P_{\mbox{\scriptsize stick}}(E_{\mbox{\scriptsize c.m.}})$ as well 
as with information of the amalgamated system, with which we can start 
to solve
a Langevin equation for 
shape evolutions and then, can obtain formation probability
$P_{\mbox{\scriptsize form}}$.    This 
means that we treat the two-body collision processes and shape 
evolutions  of the united system consistently.

\section{Realistic Calculations of $P_{\mbox{\scriptsize form}}$ and Fusion Cross sections}

   In order to describe shape evolutions starting from the pear-shape 
configuration of the amalgamated system to the spherical shape, at least, 
three parameters, say, distance between two mass centers $R$, mass asymmetry 
$\alpha$, and neck parameter $\epsilon$ in the Two-Center Parameterization\cite{bib23}.   
With OBM\cite{bib18}, the friction for the neck 
degree of the freedom is much stronger than the others and thus its motion is 
considered to be much slower than the other twos.  Then, we expect that two 
variables could describe the formation dynamics reasonably well, with the neck 
parameter $\epsilon$ freezed.    We again employ a classical dissipation-
fluctuation description, though quantum effects, such as a tunneling effect, 
might play a significant role in passing over the saddle.  

A multi-dimensional 
Langevin equation is written as usual\cite{bib24}, 

\begin{eqnarray}
\frac{dq_i}{dt}&=&(m^{-1})_{ij}\cdot p_j\\\nonumber
\frac{dp_i}{dt}&=-&\frac{\partial U}{\partial
q_i}-\frac12\frac{\partial}{\partial q_i}(m^{-1})_{jk}\cdot
p_j\cdot p_{_{\mbox{\scriptsize R}}}-\gamma_{ij}\cdot(m^{-1})_{jk}\cdot p_k+g_{ij}\cdot\Gamma_j,
\end{eqnarray}

\noindent
where $q_i$, $i$ being 1 or 2, specifies $R$, or $\alpha$, and summations are 
implicitly assumed over the repeated indices.  The inertia mass tensor $m_{ij}$ is 
calculated by Werner-Wheeler approximation\cite{bib25} and the friction tensor $\gamma_{ij}$ by 
OBM as functions of the variables $R$ and $\alpha$.  The potential $U$
is given also by 
the macroscopic LDM energy.   In case of a finite angular momentum, it should 
include the rotational energy calculated with the rigid moment of inertia.  
The random force in the r.h.s. of Eq.~(21)
is assumed to be Gaussian and is expressed with a Gaussian number 
$\Gamma_i$ and a strength tensor $g_{ij}$ which are assumed to satisfy the following 
properties,

\begin{eqnarray}
\left<\Gamma_i(t)\right>&=&0\\\nonumber
\left<\Gamma_i(t)\cdot\Gamma_j(t')\right>&=&2\cdot\delta_{ij}\cdot\delta(t-t')c\\\nonumber
g_{ik}g_{jk}&=&\gamma_{ij}\cdot T,
\end{eqnarray}

\begin{figure}
\begin{center}
\includegraphics[height=10cm]{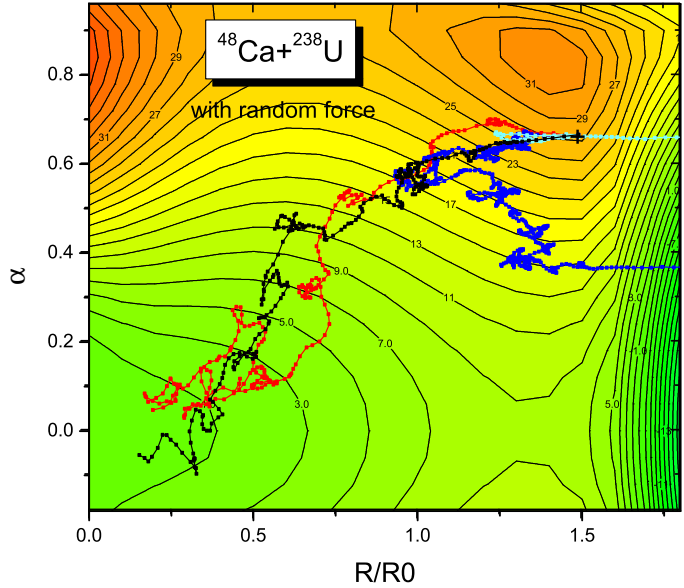}
\end{center}
\caption{Examples of trajectories in two-dimensional space of the
relative distance $R$ divided by the radius of the sphere of the total 
system $R_0$ and the mass-asymmetry $\alpha$ for
$^{48}$Ca+$^{238}$U system.
Initial momenta are taken to be zero, but Langevin forces are
calculated with the temperature given by the excitation energy of
70MeV. }
\end{figure}

\noindent
where $\left< \right>$ denotes an average over all the possible realizations.  The last 
equation expresses the dissipation-fluctuation theorem.  In order to obtain a 
formation probability, i.e., a probability for the system to overcome the 
conditional saddle point or the ridge line, we have to calculate a large 
number of trajectories.   

Examples are shown in Fig.~2, for
$^{48}$Ca-$^{238}$U system 
with zero initial radial momentum but with the temperature
corresponding to the excitation energy 70MeV, starting at the contact 
configuration.  It is seen that some trajectories go into the spherical 
configuration and its around, while the others go back to re-separation.   The 
formers consist a formation probability, while the latters do quasi-fission 
components which are to be carefully analysed in a future, including 
deformations of nascent fragments, mass drifts etc. before scission.

   Formation probabilities calculated with $\epsilon$ being 0.8 are
shown in the upper panel of Fig.~3, for $^{48}$Ca-$^{238}$U system.
And fusion probabilities 
calculated by Eq.~(2)%
, 
i.e., obtained by combining with the sticking probabilities obtained
by SFM, are shown in the lower panel of Fig. 3 for the total angular
momenta of the  
system $J=0$ and 30.
Then, 
excitation functions of fusion cross sections are calculated 
according to the following formula,

\begin{figure}
\begin{center}
\includegraphics[height=8.5cm]{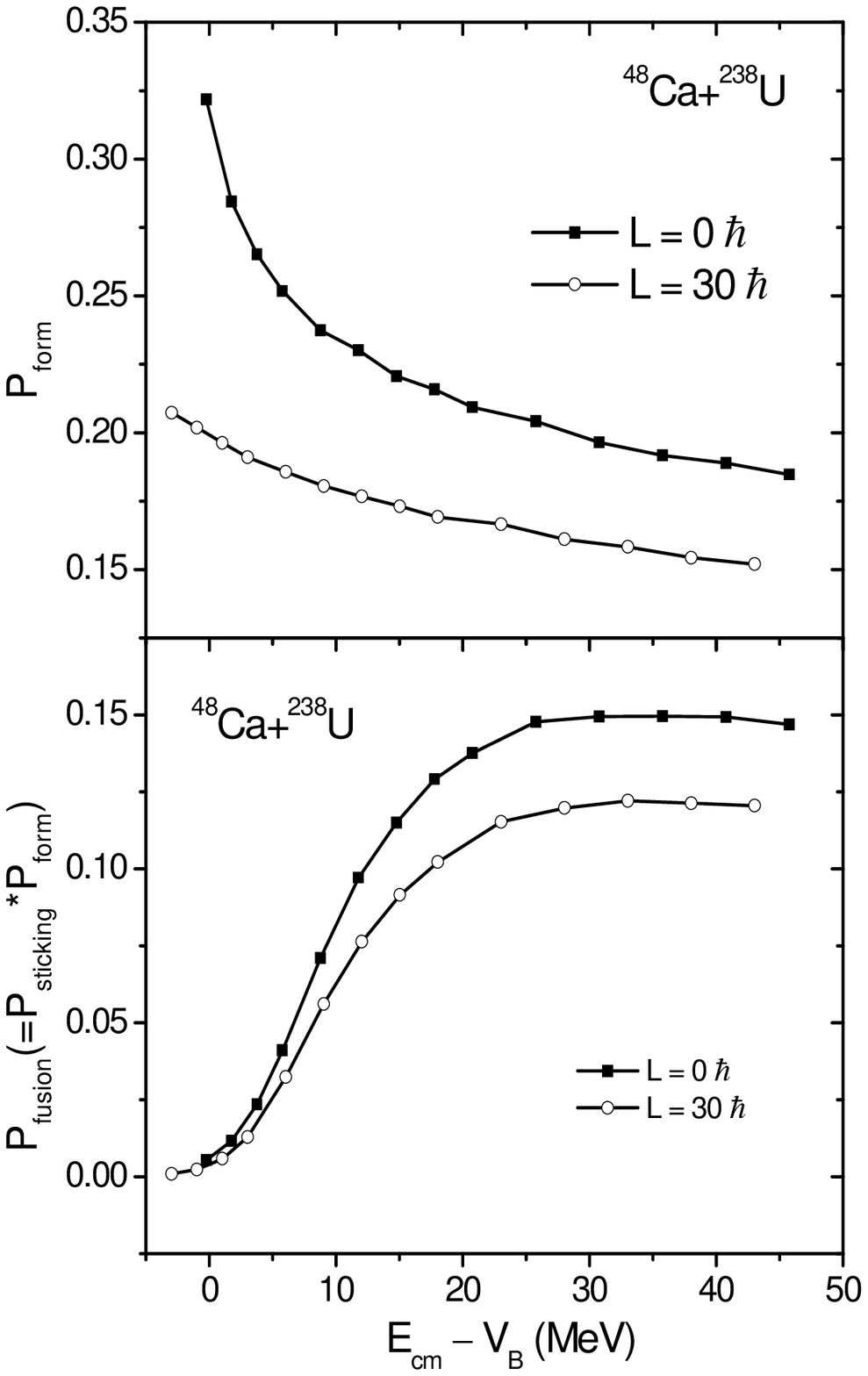}
\end{center}
\caption{The upper panel shows the formation probability
$P_{\mbox{\scriptsize form}}$ calculated for the total spin $J=0$ and
30 cases of $^{48}$Ca+$^{238}$U
system, as a function of incident energy relative to the Coulomb
barrier. 
The lower panel shows the corresponding fusion probability
$P_{\mbox{\scriptsize fusion}}$ calculated together with
$P_{\mbox{\scriptsize stick}}$ by SFM. }
\end{figure}

\begin{equation}
\sigma_{\mbox{\scriptsize fusion}}=\pi{\lb{\lambda}}^2\Sigma(2J+1)\cdot P_{\mbox{\scriptsize fusion}}.
\end{equation}

\begin{figure}
\begin{center}
\includegraphics[height=6.5cm]{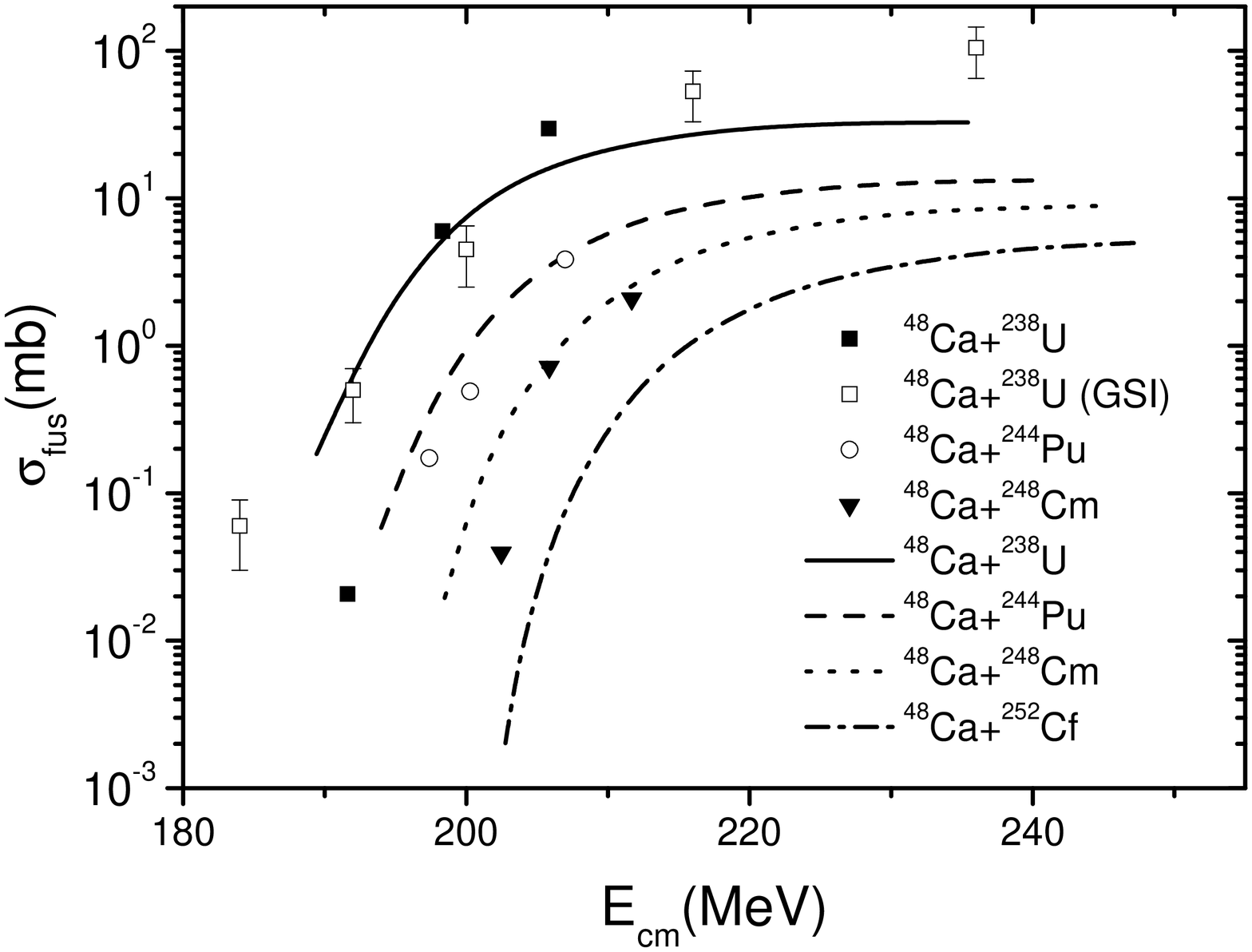}
\end{center}
\caption{Calculated fusion excitation functions are shown for
$^{48}$Ca+$^{238}$U, +$^{244}$Pu, +$^{248}$Cm, and +$^{252}$Cf
systems, together with the available experimental data from
GSI\cite{bib26} and Dubna\cite{bib27}.}
\end{figure}

\noindent
The results for $^{48}$Ca + actinide target systems are 
shown in Fig. 4, compared with the available experimental data
obtained at GSI\cite{bib26} 
and Dubna\cite{bib27}.   It is surprising that the calculations
reproduce the experimental 
data very well, not only their absolute values, but also their energy 
dependence, systematically over three systems.   A prediction is made
for $^{252}C_f$ 
target case, which should be verified by experiment.

\section{Survival Probability}

   The survival probability $P_{\mbox{\scriptsize surv}}$ is a probability for the compound system to 
survive against fission decay and charged particle emission.  
We first discuss the characteristic features qualitatively and then
present examples of realistic calculations made by a new statistical
code KEWPIE\cite{bib17} in the next section.
Since decay 
widths for the latters are small compared with those for the former and 
neutron emission, the total decay width is approximately given by
$\Gamma_f+\Gamma_n$, and then the survival probability is given
approximately by  

\begin{equation}
P_{\mbox{\scriptsize surv}}\cong\Gamma_n/\left(\Gamma_f+\Gamma_n\right)\cong\Gamma_n/\Gamma_f,
\end{equation}

\noindent
where $\Gamma_f$ and $\Gamma_n$ denote fission decay and neutron
emission widths, 
given by Bohr-Wheeler\cite{bib28} and Weisskopf\cite{bib29} formulae,
respectively.  Of course, if 
intrinsic excitation energy $E^*$ is large enough for emissions of
more than one 
neutron, the expression of Eq.~(24) 
is repeatedly used in multiplication.   
In SHE, $\Gamma_f\gg\Gamma_n$, so the second equation approximately
holds in superheavy nuclei generally except cases with very
large shell correction energies, which is 
easily seen by their approximate expressions, 

\begin{equation}
\Gamma_n\simeq e^{-B_n/T}, \;\;\; \Gamma_f\simeq e^{-B_f/T},
\end{equation}

\noindent
then the probability is given by 

\begin{equation}
P_{\mbox{\scriptsize surv}}\simeq e^{-\left(B_n-B_f\right)/T},
\end{equation}

\noindent
where $B_f$ and $B_n$ denote fission barrier height and neutron separation energy, 
respectively.  And $B_f$ is almost equal to minus of the shell correction energy, 
because macroscopic fission barriers, i.e., LDM fission barrier
$B_f^{\mbox{\scriptsize LDM}}$ is very 
small and is nearly equal to zero for SHE, due to the fact that the fissility 
parameter $x_f$ is close to 1.  

It is worth to consider how an excitation-energy dependence of the
shell correction energy comes into play.
As is expected, absolute values of the shell correction energy are reduced by 
excitation, so in the beginning of decay process.  This is well taken into 
account by Ignatyuk's prescription\cite{bib30} of excitation-energy dependence of the 
level density parameter of the spherical shape, i.e., for neutron emission,

\begin{eqnarray}
a_n&=&\bar a_n\cdot\left[1+f(E^*)\cdot\delta E/E^*\right],\\\nonumber
f(E)&=&1-\exp\left[-E^*/E_d\right],
\end{eqnarray}

\noindent
where $\bar a_n$ is an asymptotic level density parameter in high excitation, and $E^*$ 
intrinsic excitation energy of the compound nucleus.  $\delta E$ and $E_d$ denote 
shell correction energy of the ground state and so-called shell damping 
parameter, respectively.  The parameter $E_d$ is obtained to be about 18 MeV by 
calculating excitation energy dependence of the free energy with a
single particle model.  	With Eq.~(27)%
, the fission width is approximately given as follows,

\begin{eqnarray}
\Gamma_f&\cong& e^{-B_{\mbox{\scriptsize eff}}/T},\\\nonumber
B_{\mbox{\scriptsize eff}}&=&B_f+f(E^*)\cdot\delta E.
\end{eqnarray}

Then, asymptotic behaviors for $E^*\ll E_d$ and $E^*\gg E_d$ become as
follows respectively,

\begin{eqnarray}
B_{\mbox{\scriptsize eff}}\cong& B_f+E^*\cdot\delta E/E_d\rightarrow
B_f, \; &E^*\ll E_d,\\\nonumber
\cong&B_f+\delta E\rightarrow B_f^{\mbox{\scriptsize LDM}}, \;\;\;\;\;\;\;\;\; &E^*\gg E_d,
\end{eqnarray}

\noindent
where $B_f=B_f^{\mbox{\scriptsize LDM}}-\delta E$ denotes the fission
barrier height of the ground state.
As is seen from the above arguments, the survival probability $P_{\mbox{\scriptsize surv}}$ is 
crucially determined by absolute values of the shell correction energy!!
Remaining ambiguities are Kramers\cite{bib11,bib15} and collective
enhancement\cite{bib16} factors.  The  
former takes into account an effect of friction force acting on the 
fissioning degree of freedom, and is given by 

\begin{equation}
K_f=\sqrt{x^2+1}-x.
\end{equation}

\noindent
This is always smaller than 
1 and is approximately equal to $1/x$ in case of large $x$.  The collective 
enhancement factor takes into account a difference between collective level 
densities at the spherical shape and the saddle point shape.  Since the saddle 
point shape of SHE is determined by shape dependence of the shell correction 
energy, no simple formula is available.  
It should be worth to notice here that so-called Strutinski 
correction factor\cite{bib31} for Bohr-Wheeler formula
$\hbar\cdot\omega/T$ can  
be considered to be a part of the collective enhancement factor, i.e., the 
part from the fissioning collective degree of freedom, though the main 
part of the enhancement is expected to be that of the rotational
degrees of freedom.

\section{Preliminary Results of Residue Cross Sections}

In order to make realistic calculation of the survival probability we
have made a new statistical code KEWPIE (Kyoto Evaporation Width
calculation Program with tIme Evolution)\cite{bib17}. 
This program treats both the production of residue as a function of the
time and the final residue production. In the present case we will
only consider the amount of nuclei remaining at the end of the
disintegration cascade.  Detailed formalism and  the computer code will
be published elsewhere. 

This program includes the main features required in the section
6. However  in this code the evaporation width of particle is
calculated more accurately in the Hauser-Feshbach formalism
\cite{bib32}.  
Moreover the evaporation of protons, alphas and gammas  are included
in the program.  Calculation of  fission width is done according to
Bohr-Wheeler formula with the transmission coefficient by Hill and
Wheeler\cite{bib33} and with Strutinski correction factor.  
The fission barrier $B_f^{LDM} $ is that of the
empirical formula given  for heavy elements by K.H. Schmidt {\it et
al.} in reference \cite{bib34}.

The level density parameters $\bar a_n$ and $a_f$ are calculated with
T\"oke and
Swiatecki formula\cite{bib35} by taking into account of shapes of the
ground state and the saddle point.
At the ground state the shape of the nucleus is assume to be spherical
and we take into account the shell correction effect with the Ignatyuk
prescription \cite{bib30} with $E_d = 18 $ MeV as given in Eq.~(27). 
At saddle point, deformation is evaluated by the Hasse and Myers formula
\cite{bib36} and no shell correction effect are taken into account. 

\begin{figure}
\begin{center}
\rotatebox{270}{\includegraphics[height=6.5cm]{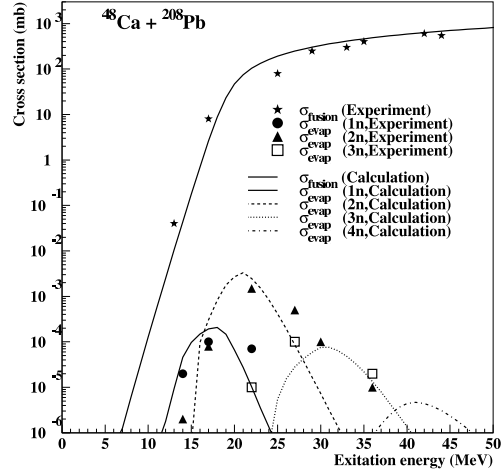}}
\end{center}
\caption{Residue cross sections calculated by the statistical code
KEWPIE are shown for $^{48}$Ca+$^{208}$Pb, compared with the
experimental data\cite{bib38}.}
\end{figure}

\begin{figure}
\begin{center}
\rotatebox{270}{\includegraphics[height=6.5cm]{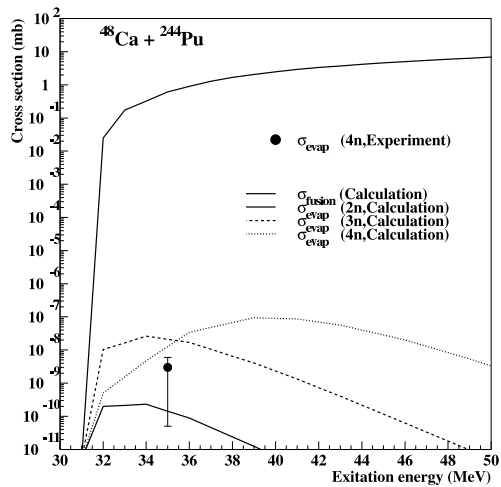}}
\end{center}
\caption{Calculated residue cross sections are shown for
$^{48}$Ca+$^{244}$Pu system, together with Dubna data\cite{bib39}.
Details on the parameters are given in the text.}
\end{figure}

The  KEWPIE calculation  has few free parameters, the scaling factor
of the shell correction taken from M\o ller et al.'s table\cite{bib2} and the
parameters of  Kramer factor $K_f$. 
The latter is calculated with $\hbar \omega = 1$ MeV and a friction
factor $\beta = 5 \times 10^{20} sec^{-1}$.

For $^{48}$Ca+$^{208}$Pb system, fusion probabilities are calculated
with the proximity 
potential\cite{bib37}, because no fusion hindrance is observed there.  
With the parameters fixed, we calculate $xn$ residue cross sectoins,  
whose results are shown
in   the Fig.~5.
The experimental cross sections\cite{bib38} are seen to be well
reproduced, which
appears to guarantee the code KEWPIE.
 
For $^{48}$Ca+$^{244}Pu$  reaction, 
we use fusion probabilities calculated according to Eq.~(2) with the
realistic calculations of $P_{\mbox{\scriptsize form}}$ given in
section 5.
As discussed in section 6, the crucial parameter in the survival
probability $P_{\mbox{\scriptsize surv}}$ is the shell correction
energy. 
The scaling factor of
$2/3$ or even smaller has turned out to be necessary to the shell
correction energies 
of P. M\o ller et al. in order to be 
consistent with the data\cite{bib39}, as is shown on Fig.~6.
As  theoretical values of the shell correction energy are very  different
from one model to another, more precise investigations are desired. 
We are now studying the reactions of $^{48}$Ca + actinide targets,
using several predictions of
the shell correction energy, which could be very
informative on the models of nuclear structure for heavy and
superheavy nuclei. 

\section*{Acknowldgements}
Y. Abe acknowledges long-standing fruitful collaborations 
with T. Wada, D. Boilley, C.W. Shen, G. Kosenko and B. Giraud, with
the results of which the present contribution is mostly written.
B. Bouriquet thanks the supports for the post-doctoral position
provided by JSPS which gives him an opportunity to work at Yukawa
Institute for Theoretical Physics, Kyoto University.  
This work is partially supported by the Grant-in-Aids of JSPS
(no.~1340278).

\end{document}